\documentclass[%
 reprint,
showpacs,preprintnumbers,
 amsmath,amssymb,
 aps,
]{revtex4-1}

\usepackage{graphicx}
\usepackage{dcolumn}
\usepackage{bm}

\begin{document}


\title{Cho decomposition, Abelian gauge fixing and monopoles in G($2$) Yang-Mills theory}

\author{Z. Dehghan}
\email{zeinab.dehghan@ut.ac.ir}
\author{S. Deldar}%
 \email{sdeldar@ut.ac.ir}
\affiliation{%
Department of Physics, University of Tehran, P. O. Box 14395/547, Tehran 1439955961, Iran
}%

\date{\today}

\begin{abstract}

By extending the Cho decomposition method to G($2$) gauge group, monopoles of this group are studied. Since SU($2$) and SU($3$) are subgroups of G($2$), discussions are done mostly based on these subgroups of G($2$). A direct relation between root vectors of G($2$) and the associated magnetic charges is presented by group theoretical issues. In addition, G($2$) monopoles are obtained by an Abelian gauge fixing method, and it is shown that the results agree with the ones we obtain by the Cho decomposition method.

\end{abstract}

\pacs{11.15.Ha, 12.38.Aw, 12.38.Lg, 12.39.Pn}
\maketitle


\section{Introduction}\label{sec1}

Monopoles are among the most popular candidates for describing quark confinement. Therefore, studying their properties in various gauge groups helps to understand this phenomenon. G($2$), as the smallest simply connected exceptional group which contains both SU($2$) and SU($3$) gauge groups as its subgroups, is an appropriate group to be studied in this respect. Lattice calculations \cite{confinement1, confinement2, well1, well2, confinement3} confirmed the confinement of color sources at intermediate distances for the G($2$) gauge group. There are also some phenomenological discussions and results based on the thick center vortex model \cite{vortex1, deldar1, deldar2, deldar3}. In addition to these articles where people studied the confinement properties by vortex configurations, some other studies based on monopole properties discussed the confinement mechanism \cite{shnir, kondo}. In this article, using Cho decomposition and Abelian gauge fixing methods, we study monopoles of G($2$) and some of their properties, mainly based on its SU($2$) and SU($3$) subgroups.	

The decomposition method introduced by Faddeev-Niemi  \cite{fn1, fn2, fn3} and the one represented by Cho \cite{chord211980, chord231981} are two of the methods which can be applied to decompose the Yang-Mills fields of the non-Abelian gauge groups. In these two  methods, Yang-Mills fields are decomposed in terms of the  variables which are more appropriate for describing the theory in the low energy limit and the confinement regime. In this paper, we extend the Cho decomposition of SU($2$) and SU($3$) gauge groups to the non-Abelian G($2$) gauge group. We show that by Cho decomposition, one can study topologically properties of G($2$) gauge group and obtain Wu-Yang monopoles. In addition, a restricted Lagrangian for the confinement regime can be obtained. Finally, by Abelian gauge fixing, we find monopoles of G($2$). The gluon field acquires a singularity in the vicinity of the points in space where Abelian gauge fixing fails and magnetic monopoles are formed. Indeed, there are some points in the space where the Abelian gauge fixing becomes ill-defined. Topological defects such as magnetic monopoles \cite{Suganuma, Ichie, pepe, ripka} are extracted from these points, and some properties of them like their charges can be obtained by this method.

In Sec. II, a brief review on G($2$) gauge group and its properties are given. Cho decomposition of G($2$) is discussed in Sec. III. Then, using this decomposition, we find monopoles and the magnetic charges of G($2$) gauge group by its SU($2$) subgroups. The generalized quantization condition of G($2$), which  contains two integers and is determined as a consequence of having two different types of monopoles, is calculated in this section as well. In Sec. IV, we extract monopoles of G($2$) by Abelian gauge fixing method. Finally, the summary and discussions are given in Sec. V.

\section{Some properties of G($2$) gauge group}\label{sec22}

In this section, we review some basic properties of G($2$) gauge group. G($2$) is the smallest simply connected exceptional group which is its own universal covering group with only a trivial center \cite{pepe1,pepe2,well1}. It has 14 generators and since the rank of the group is 2, two of the generators are simultaneously diagonal. It is a subgroup of the real group $SO(7)$ with rank 3 and 21 generators, and therefore, it is  a real group. The determinant of the $7\times7$ real orthogonal matrices $U$ of the group $SO(7)$ is 1, and
\begin{equation}\label{e1}
U U^{\dagger}=1.
\end{equation}
In addition, G($2$) elements satisfy a constraint called the cubic constraint,
\begin{equation}\label{e2}
T_{abc}=T_{def} U_{da} U_{eb} U_{fc},
\end{equation}
where $T$ is a totally antisymmetric tensor, and its nonzero elements are
\begin{equation}\label{e3}
\hspace{-0.5mm}T_{127}=\hspace{-0.5mm}T_{154}=\hspace{-0.5mm}T_{163}=\hspace{-0.5mm}T_{235}=\hspace{-0.5mm}T_{264}=\hspace{-0.5mm}T_{374}=\hspace{-0.5mm}T_{576}=\hspace{-0.5mm}1.
\end{equation}
As a result of the above constraints, the number of generators of G($2$) is reduced to $14$. The dimensions of the fundamental and adjoint representations of G($2$) are $7$ and $14$, respectively. Since all G($2$) representations are real,  representation $\{7\}$ is equivalent to its complex conjugate, and therefore, G($2$) ``quarks" and ``antiquarks" are indistinguishable.

SU($2$) and SU($3$) are subgroups of the G($2$) gauge group. Since G($2$) is rank 2 and SU($3$) as its subgroup has the same rank, some properties of G($2$) are the same as SU($3$). For example, the second and the third homotopy groups of G($2$) include monopoles and instantons \cite{instanton1, instanton2, instanton3}, respectively, the same as SU($3$). The second homotopy group of G($2$) is
\begin{equation}\label{e9}
\Pi_2[G(2)/U(1)^2]=\Pi_1[U(1)^2]=\mathbb{Z}^2.
\end{equation}
It means that when G($2$) group is broken globally to its Cartan subgroup $U(1)^2$, two types of monopoles appear. Extracting these monopoles by decomposition methods is one of the goals of this paper. The third homotopy group is
\begin{equation}\label{e8}
\Pi_3[G(2)]=\mathbb{Z}.
\end{equation}
It implies that instantons are present in G($2$) theory. The first homotopy group of G($2$) and SU($3$) are
\begin{align}\label{e10}
\begin{aligned}
&\Pi_1[G(2)/I]=\Pi_1[G(2)]=0,\\
&\Pi_1[SU(3)/\mathbb{Z}_3]=\mathbb{Z}_3,
\end{aligned}
\end{align}
where $I$ and $\mathbb{Z}_3$ are the center groups of G($2$) and SU($3$), respectively. In contrast to SU($3$), G($2$) has no center vortices because it has only a trivial center.
 
Under SU($3$) subgroup transformations, the seven-dimensional fundamental representation of G($2$) is decomposed into the SU($3$) fundamental representations,
\begin{equation}\label{e4}
\{7\}=\{3\}\oplus\{\bar{3}\}\oplus\{1\}.
\end{equation}
In other words, a G($2$) ``quark" $\{7\}$ contains an SU($3$) quark $\{3\}$, an SU($3$) antiquark $\{\bar{3}\}$, and an SU($3$) singlet $\{1\}$. We recall that the $\{3\}\oplus\{\bar{3}\}$ contained in the $\{7\}$th dimensional representation of G($2$) corresponds to a real reducible six-dimensional representation of SU($3$).

G($2$) contains the generators of SU($3$) as its subgroup, and it is possible to choose its generators in such a way that the SU($3$) generators are clearly observed. The following choices for the eight generators of 
the seven-dimensional fundamental representation of G($2$) \cite{pepe1} are made:
\begin{equation}\label{e5}
     T_i=N_i\begin{pmatrix}
       \lambda_i && \multicolumn{2}{c}{\text{\kern-0.7em\smash{\raisebox{-1.3ex}{\mbox{\Large 0}}}}} \\
       & -\lambda^{*}_i &  \\
       \multicolumn{2}{c}{\text{\kern-1em\smash{\raisebox{-0.5ex}{\mbox{\Large 0}}}}} && 0
     \end{pmatrix},
\end{equation}
where $\lambda_i (i=1,2,...,8)$ are the $3\times3$ Gell-Mann generators of SU($3$). $N_i$ is obtained from the normalization condition. With this choice of generators, the decomposition of G($2$) to its SU($3$) subgroup brought in Eq.\;\eqref{e4}, is more understandable. The representation we choose consists of complex numbers. However, it is equivalent to a representation that is entirely real. With our choice of basis, the SU($3$) subgroup generators $T_3$ and $T_8$ are diagonal.

\begin{figure}[th]  \includegraphics[width=8.5cm]{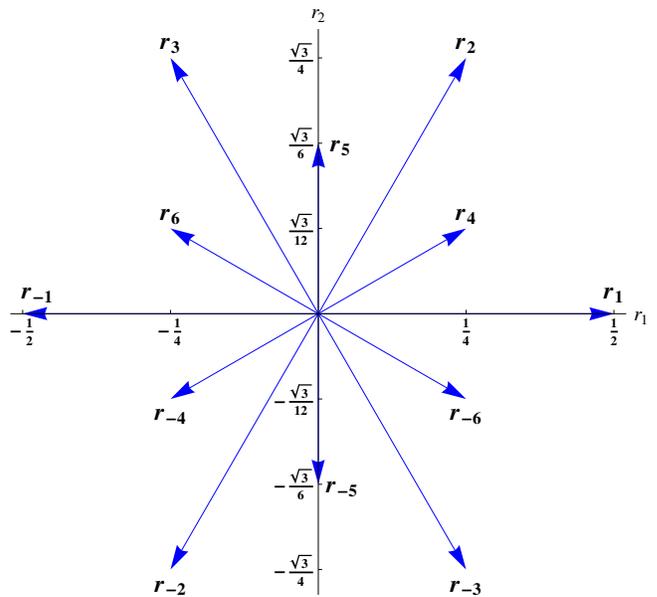}
\caption{\label{rootdiagram} \small
Root diagram of G($2$).}
\end{figure}

Since the generators of a group can be found by studying the root and weight diagrams of that group, it is possible to  find the remaining generators of G($2$) by this method. If $\mathbf{r}_\alpha$ is a root, then $\mathbf{r}_{-\alpha}\equiv -\mathbf{r}_\alpha$ is also a root.  The set of positive and negative root vectors makes the root space $R$. As mentioned before, the non-Abelian 
G($2$) gauge group has fourteen generators and the rank of this group is two; therefore, it has twelve roots $\mathbf{r}_\alpha (\alpha=\pm1,\pm2, ...,\pm6\in R)$. The root diagram of G($2$) \cite{modern} is shown in Fig.\;(\ref{rootdiagram}). Since the rank of G($2$) is $2$, the root diagram is two dimensional and the vertical and horizontal axes are shown by $r_{2}$ and $r_{1}$, respectively. $\mathbf{r}_{1}$ and $\mathbf{r}_{4}$ are two of the simple roots of G($2$) where the ratio of the lengths of these two vectors are $\sqrt{3}$ and the angle between them is ${\pi}/{6}$. Simple roots are the roots which are linearly independent, and other roots can be constructed from them. The following normalization condition is chosen: 
\begin{equation}\label{normal}
\sum_{\alpha} r_{\alpha}^j r_{\alpha}^k=\delta^{jk},\quad(j,k=1,2),
\end{equation}
where $r_{\alpha}^k$ is the $k$th component of the root $\mathbf{r}_{\alpha}$. G($2$) contains two Carton generators $H_k(k=1,2)$ and twelve stepping generators $E_{\alpha} (\alpha\in R)$, where they satisfy the standard form of the commutation relations \cite{modern, grouptheory1},
\begin{align}\label{e6}
\begin{aligned}
&[H_j,H_k]=0,\\
&[H_k,E_{\alpha}]=r_{\alpha}^k E_{\alpha},\\
&[E_{\alpha},E_{\beta}]=\begin{cases}
N_{\alpha,\beta} E_{\alpha+\beta} \quad &(\alpha+\beta\in R)\\
r_{\alpha}^k H_k \quad &(\alpha+\beta=0).\\
0\quad &(\mbox{otherwise})
\end{cases}
\end{aligned}
\end{align}
$N_{\alpha,\beta}$ are real constants and
\begin{equation}\label{e7}
E_{-\alpha}=E^{\dagger}_{\alpha}, \quad H_{k}=H^{\dagger}_{k}.
\end{equation}
The non-Hermitian generators $E_{\alpha}$ and $E_{-\alpha}$ are proportional to the raising and lowering operators. There is a one-to-one correspondence between the generators $E_{\alpha}$ and the roots $\mathbf{r}_{\alpha}$. $H_k$'s are sometimes called the generators corresponding to zero roots.

From Eq.\;\eqref{e6}, $[H_1,H_2]=0$; therefore, it is possible to diagonalize these two matrices simultaneously. For any representation of G($2$), for diagonalized generators $H_k (k=1,2)$,  the following eigenvalue equation is applied: 
 \begin{equation}\label{hk} 
  H_k\rvert\mathbf{m}_{a}\rangle=m^{k}_{a}\rvert\mathbf{m}_{a}\rangle, \quad (a=1,2,...,d_{IR}).
 \end{equation}
The vector $\mathbf{m}_{a}$ is called the weight, and $m^{k}_{a}$ is the $k$th component of the weight $\mathbf{m}_{a}$. The eigenstates of any representation of G($2$) are shown by $\rvert\mathbf{m}_{a}\rangle$, which are defined by the weight vectors. The number of weights is denoted by $d_{IR}$---the dimension of the representation. Weight diagrams are constructed by weight vectors. The generators of the fundamental representation of G($2$) are found by using root and weight diagrams in some papers \cite{pepe1,modern}. In the Appendix, we bring another form of the G($2$) generators. Indeed, rather than working with real generators $H_k$ and $E_{\alpha}$, we prefer to work with $T_i (i=1,2,...,14)$ generators: $T_3=H_1$, $T_8=H_2$ and other $T_i$'s which are complex combinations of generators $E_{\alpha}$'s \eqref{ap5}. The first eight generators of $T_i$ have the typical form of Eq.\;\eqref{e5}. We show that from this form of $T_i$'s, we can learn many interesting properties of SU($2$) and SU($3$) subgroups of G($2$).

\section{Cho decomposition of G($2$)}\label{sec2}

The nonzero second homotopy group of G($2$) discussed in the last section, confirms the existence of monopoles. There are some various methods for extracting topological structures such as monopoles. Cho decomposition is one of these methods in which the Yang-Mills fields are decomposed in terms of variables that are more appropriate for describing the theory in the confinement regime. Since the theory is restricted, the dynamical degrees of freedom are reduced, and a self-consistent and nontrivial subset of the original gauge theory is provided. As a result of this decomposition, the original Yang-Mills theory turns into electrodynamics with magnetic monopoles. Cho decomposition and Abelian gauge fixing are the two methods that we use and apply to G($2$) gauge group to find monopoles and their properties in this article. 

First, we review Cho decomposition for the simplest gauge group, SU($2$) \cite{chord211980}. We assume that vectors $\hat{n}_1,\hat{n}_2,\hat{n}_3=\hat{n}$ are the local orthonormal SU($2$) basis, where $\hat{n}$ is an isotriple vector field that gives the Abelian direction at each space-time point. A Cho decomposition for SU($2$) Yang-Mills field $\mathbf{A}_{\mu}$, is obtained by assuming that in the low energy limit,
\begin{equation}\label{q1}
D_{\mu}\hat{n}=(\partial_{\mu}+g\mathbf{A}_{\mu}\times)\hat{n}=0,
\end{equation}
where $g$ is the Yang-Mills coupling constant. The condition \eqref{q1} makes a restriction on $\mathbf{A}_{\mu}$. We characterize the restricted Yang-Mills field by $\hat{A}_{\mu}$. The condition of Eq. \eqref{q1} indicates that the restricted Yang-Mills field is a field which leaves $\hat{n}$ invariant under parallel transport. $\hat{A}_{\mu}$ is obtained by solving Eq.\;\eqref{q1},
\begin{equation}\label{q2}
\mathbf{A}_{\mu} \longrightarrow \hat{A}_{\mu}=A_{\mu}\hat{n}-\dfrac{1}{g}\hat{n}\times \partial_{\mu} \hat{n},\quad A_{\mu}=\hat{n}\cdot\mathbf{A}_{\mu},
\end{equation}
where $A_{\mu}$ is the Abelian component of the potential that is not restricted by Eq.\;\eqref{q1}. The restricted potential \eqref{q2} contains two parts: a nontopological unrestricted part $A_{\mu}$ which is called the electric potential and a topological restricted part which is related to the magnetic potential. Therefore in the infrared limit, $\hat{A}_{\mu}$ dominates and contains electric and magnetic potentials in a dual symmetric way. Using the restricted potential of Eq.\;\eqref{q2}, the field strength $\hat{G}_{\mu\nu}$ is given by
\begin{align}\label{c1}
\begin{aligned}
\hat{G}_{\mu\nu}&=\partial_{\mu}\hat{A}_{\nu}-\partial_{\nu}\hat{A}_{\mu}+g\hat{A}_{\mu}\times\hat{A}_{\nu}\\
&=(F_{\mu\nu}+H_{\mu\nu})\hat{n},
\end{aligned}
\end{align}
where
\begin{equation}\label{c2}
F_{\mu\nu}=\partial_{\mu}A_{\nu}-\partial_{\nu}A_{\mu}, \quad H_{\mu\nu}=-\dfrac{1}{g}\hat{n}\cdot(\partial_{\mu}\hat{n}\times\partial_{\nu}\hat{n}).
\end{equation}
Obviously, the field strength $\hat{G}_{\mu\nu}$, has a dual structure since it can be decomposed into an electric field strength $F_{\mu\nu}$ and a magnetic field strength $H_{\mu\nu}$. In Eq.\;\eqref{c2}, $F_{\mu\nu}$ is expressed in terms of the electric potential $A_{\mu}$. Correspondingly, we can define a magnetic potential $C_{\mu}$ and express $H_{\mu\nu}$ in terms of this potential
\begin{equation}\label{c3}
H_{\mu\nu}=-\dfrac{1}{g}\hat{n}\cdot(\partial_{\mu}\hat{n}\times\partial_{\nu}\hat{n})=\partial_{\mu}C_{\nu}-\partial_{\nu}C_{\mu}.
\end{equation} 
Indeed, $A_{\mu}$ and $C_{\mu}$ are electric and magnetic contributions of the gluon field $\hat{A}_{\mu}$, respectively. The singularity observed as a monopole in the color direction $\hat{n}$ is expected from the statement $\Pi_2(SU(2)/U(1))=\Pi_2(S^2)=\mathbb{Z}$. We choose $A_{\mu}=0$ and a hedgehog form for $\hat{n}$ (color direction correlated with configuration direction),
\begin{equation}\label{c4}
\hat{n}=\hat{n}_3=\hat{r}=\left(\begin{matrix}
\sin\theta\cos\varphi\\
\sin\theta\sin\varphi\\
\cos\varphi
\end{matrix}\right).
\end{equation}
From Eq.\;\eqref{c3}, the magnetic field strength is given by
\begin{equation}\label{c5}
H_{\mu\nu}=-\dfrac{1}{g}\sin\theta(\partial_{\mu}\theta\partial_{\nu}\varphi-\partial_{\nu}\theta\partial_{\mu}\varphi)=\partial_{\mu}C_{\nu}-\partial_{\nu}C_{\mu}.
\end{equation}
We can consider $C_{\mu}=\dfrac{1}{g}(1+\cos\theta)\partial_{\mu}\varphi$ \cite{oxman2008} as an answer for Eq.\;\eqref{c5}, where the components of this magnetic potential are
\begin{equation}\label{c6}
C_{\theta}=C_{r}=0,\quad C_{\varphi}=\dfrac{1}{g}\dfrac{(1+\cos\theta)}{r\sin\theta}.
\end{equation}
Therefore, the magnetic potential has a singularity at $\theta=0$. In $4$D, this magnetic potential describes a static Wu-Yang monopole located at the origin with a Dirac string along the positive $z$ axis. Since the system is Abelianized (diagonalized) in the third axis direction in color space ($\hat{n}=\hat{n}_3$), we can write the singular part of the gluon field as the following:
\begin{equation}\label{c7}
\hat{A}=\hat{A}_a T_a \longrightarrow \hat{A}^{S}=\mathbf{C} T_{d}=\dfrac{1}{g}\dfrac{(1+\cos\theta)}{r\sin\theta}(\dfrac{\sigma_{3}}{2})\hat{\varphi},
\end{equation}
where a sum on ``$a=1,2,3$" in color space is denoted and $T_{d}$ indicates the diagonal generator of SU($2$), which is $\sigma_{3}/2$. Therefore, the magnetic charge associated to the pointlike monopole in SU($2$) is
\begin{equation}\label{charge}
g_{m}=\dfrac{4\pi}{g}(\dfrac{\sigma_{3}}{2}),
\end{equation}
which corresponds to the diagonal generator of the group. We recall that choosing $\hat{n}=\pm\hat{r}$ gives monopole (antimonopole).

Later, Cho also applied this decomposition to SU($3$) Yang-Mills fields \cite{cholett1980,cho2014}. In this section, we extend this method to decompose the Yang-Mills field of G($2$) gauge group. We consider the direction vectors $\hat{n}_i (i=1,2,...,14)$ as the local orthonormal G($2$) basis corresponding to the generators $T_i$'s brought in the Appendix. The Abelian directions are assumed to be $\hat{n}_3=\hat{n}$ and $\hat{n}_8=\hat{n}'$ corresponding to the diagonal generators $T_3$ and $T_8$, respectively. These two Abelian directions are represented by a two-dimensional vector $\mathbf{n}$,
\begin{equation}\label{s2eq2}
\mathbf{n}=(\hat{n},\hat{n}').
\end{equation}
Since G($2$) contains two Abelian directions, Cho condition is given by
\begin{equation}\label{s2eq3}
D_{\mu}\hat{n}=0, 
\quad D_{\mu}\hat{n}'=0.
\end{equation}
Indeed, the first condition automatically guarantees the second condition and $\hat{n}'\propto\hat{n}\ast\hat{n}$, where the star denotes the symmetric d product \cite{cholett1980}. By solving Eq.\;\eqref{s2eq3}, a restricted potential for G($2$) gauge group is found,
\begin{equation}\label{s2eq4}
\hat{A}_{\mu}=A_{\mu}\hat{n}-\dfrac{1}{g}\hat{n}\times \partial_{\mu} \hat{n}+A'_{\mu}\hat{n}'-\dfrac{1}{g}\hat{n}'\times \partial_{\mu} \hat{n}',
\end{equation}
where $A_{\mu}=\hat{n}\cdot\mathbf{A}_{\mu}$ and $A'_{\mu}=\hat{n}'\cdot\mathbf{A}_{\mu}$ are Abelian components of the potential. It is obvious that the restricted potential contains two parts, the Maxwell part and the Dirac part. 

As we show in the Appendix \eqref{a2}, the space of G($2$) can be covered by six SU($2$) subgroups. Thus, for the corresponding local directions in the internal space, we can define the six following categories:
\begin{align}\label{s2eq6}
\begin{aligned}
&(\dfrac{1}{2}\hat{n}_1,\dfrac{1}{2}\hat{n}_2,\hat{n}^1),\quad (\dfrac{1}{2}\hat{n}_4,\dfrac{1}{2}\hat{n}_5,\hat{n}^2),\\ &(\dfrac{1}{2}\hat{n}_6,\dfrac{1}{2}\hat{n}_7,\hat{n}^3),\quad
(\dfrac{1}{2\sqrt{3}}\hat{n}_9,\dfrac{1}{2\sqrt{3}}\hat{n}_{10},\hat{n}^4),\\ &(\dfrac{1}{2\sqrt{3}}\hat{n}_{11},\dfrac{1}{2\sqrt{3}}\hat{n}_{12},\hat{n}^5),\quad (\dfrac{1}{2\sqrt{3}}\hat{n}_{13},\dfrac{1}{2\sqrt{3}}\hat{n}_{14},\hat{n}^6),
\end{aligned}
\end{align}
where $\hat{n}^p (p=1,2,...,6)$ are Abelian directions of SU($2$) subgroups, which are combinations of Abelian directions $\hat{n}$ and $\hat{n}'$ in G($2$),
\begin{align}\label{s2eq7}
\begin{aligned}
&\hat{n}^1=\dfrac{1}{2}\hat{n},\quad \hat{n}^2=\dfrac{1}{4}\hat{n}+\dfrac{\sqrt{3}}{4}\hat{n}',\quad \hat{n}^3=-\dfrac{1}{4}\hat{n}+\dfrac{\sqrt{3}}{4}\hat{n}', \\
&\hat{n}^4=\dfrac{1}{4}\hat{n}+\dfrac{\sqrt{3}}{12}\hat{n}',\quad \hat{n}^5=\dfrac{\sqrt{3}}{6}\hat{n}',\quad \hat{n}^6=-\dfrac{1}{4}\hat{n}+\dfrac{\sqrt{3}}{12}\hat{n}'.
\end{aligned}
\end{align}
Comparing Eq.\;\eqref{s2eq7} with root vectors of G($2$) in Fig.\;(\ref{rootdiagram}), we can expand $\hat{n}^p$ in terms of positive root vectors,
\begin{equation}\label{s2eq8}
\hat{n}^p=\mathbf{n}\cdot \mathbf{r}_p,
\end{equation}
where we show positive root vectors of G($2$) by $\mathbf{r}_p$. Indeed, $\mathbf{r}_p=\mathbf{r}_{\alpha>0}$. We can express the restricted potential \eqref{s2eq4} in terms of Abelian directions $\hat{n}^p$ and the corresponding Abelian potentials $A_{\mu}^p$,
\begin{align}\label{s2eq10}
\begin{aligned}
&\hat{A}_{\mu}=2\sum_{p=1}^{6}\hat{A}_{\mu}^p,\quad\hat{A}_{\mu}^p=A_{\mu}^p\hat{n}^p-\dfrac{1}{g}\hat{n}^p \times \partial_{\mu}\hat{n}^p,\\ 
&A_{\mu}^p=\tilde{A}_{\mu}\cdot \mathbf{r}_p,\quad \tilde{A}_{\mu}=(A_{\mu},A'_{\mu}).
\end{aligned}
\end{align}
We recall that the Einstein summation rule is not applied.
This representation of restricted potential has a compact form and contains six SU($2$) restricted potentials. Because of summation on $p$, they are not independent. 

Back to the potential of Eq.\;\eqref{s2eq4}, the field strength can be obtained
\begin{align}\label{s2eq11}
\begin{aligned}
\hat{G}_{\mu\nu}&=\partial_{\mu}\hat{A}_{\nu}-\partial_{\nu}\hat{A}_{\mu}+g\hat{A}_{\mu}\times\hat{A}_{\nu}\\
&=G_{\mu\nu}\hat{n}+G'_{\mu\nu}\hat{n}',
\end{aligned}
\end{align}
where
\begin{align}\label{s2eq12}
\begin{aligned}
&G_{\mu\nu}=\partial_{\mu}B_{\nu}-\partial_{\nu}B_{\mu}=F_{\mu\nu}+H_{\mu\nu},\quad B_{\mu}=A_{\mu}+C_{\mu},\\
&G'_{\mu\nu}=\partial_{\mu}B'_{\nu}-\partial_{\nu}B'_{\mu}=F'_{\mu\nu}+H'_{\mu\nu},\quad B'_{\mu}=A'_{\mu}+C'_{\mu},\\
&F_{\mu\nu}=\partial_{\mu}A_{\nu}-\partial_{\nu}A_{\mu},\quad F'_{\mu\nu}=\partial_{\mu}A'_{\nu}-\partial_{\nu}A'_{\mu},\\
&H_{\mu\nu}=-\dfrac{1}{g}\hat{n}\cdot(\partial_{\mu}\hat{n}\times\partial_{\nu}\hat{n})=\partial_{\mu}C_{\nu}-\partial_{\nu}C_{\mu},\\
&H'_{\mu\nu}=-\dfrac{1}{g}\hat{n}'\cdot(\partial_{\mu}\hat{n}'\times\partial_{\nu}\hat{n}')=\partial_{\mu}C'_{\nu}-\partial_{\nu}C'_{\mu}.
\end{aligned}
\end{align}
The field strength contains electric ($F_{\mu\nu}$, $F'_{\mu\nu}$) and magnetic ($H_{\mu\nu}$, $H'_{\mu\nu}$) parts. The electric Abelian field strengths $F_{\mu\nu}$ and $F'_{\mu\nu}$ are expressed in terms of the electric potentials $A_{\mu}$ and $A'_{\mu}$. The magnetic Abelian field strengths $H_{\mu\nu}$ and $H'_{\mu\nu}$ are expressed in terms of the magnetic potentials $C_{\mu}$ and $C'_{\mu}$, where these magnetic potentials determine the non--Abelian monopoles. This represents a duality between electric and magnetic potentials. Therefore, we can redefine $\hat{G}_{\mu\nu}$ in terms of Abelian potentials $B_{\mu}$ and $B'_{\mu}$, the dual potentials. One can also define the field strength in terms of Abelian directions of SU($2$) subgroups $\hat{n}^p$ and their corresponding Abelian field strength, $G_{\mu\nu}^p$,
\begin{align}\label{g}
\begin{aligned}
\hat{G}_{\mu\nu}&=2\sum_{p=1}^{6}\hat{G}_{\mu\nu}^p,\\ \hat{G}_{\mu\nu}^p&=G_{\mu\nu}^p \hat{n}^p=(\partial_{\mu}B_{\nu}^p-\partial_{\nu}B_{\mu}^p)\hat{n}^p\\
&=(F_{\mu\nu}^p+H_{\mu\nu}^p)\hat{n}^p=F_{\mu\nu}^p \hat{n}^p-\dfrac{1}{g}\partial_{\mu}\hat{n}^p \times \partial_{\nu}\hat{n}^p,
\end{aligned}
\end{align}
where
\begin{align}\label{cg}
\begin{aligned}
&B_{\mu}^p=A_{\mu}^p+C_{\mu}^p,\quad F_{\mu\nu}^p=\partial_{\mu}A_{\nu}^p-\partial_{\nu}A_{\mu}^p,\\
&H_{\mu\nu}^p=\partial_{\mu}C_{\nu}^p-\partial_{\nu}C_{\mu}^p=-\dfrac{1}{\hat{n}^p\cdot\hat{n}^p}\dfrac{1}{g}\hat{n}^p\cdot(\partial_{\mu}\hat{n}^p\times\partial_{\nu}\hat{n}^p),\\
&B_{\mu}^p=\tilde{B}_{\mu}\cdot \mathbf{r}_p,\quad \tilde{B}_{\mu}=(B_{\mu},B'_{\mu}),\\
&C_{\mu}^p=\tilde{C}_{\mu}\cdot \mathbf{r}_p,\quad \tilde{C}_{\mu}=(C_{\mu},C'_{\mu}),\\
&F_{\mu\nu}^p=\tilde{F}_{\mu\nu}\cdot \mathbf{r}_p,\quad \tilde{F}_{\mu\nu}=(F_{\mu\nu},F'_{\mu\nu}),\\
&H_{\mu\nu}^p=\tilde{H}_{\mu\nu}\cdot \mathbf{r}_p,\quad \tilde{H}_{\mu\nu}=(H_{\mu\nu},H'_{\mu\nu}).
\end{aligned}
\end{align}
Abelian field strength $F_{\mu\nu}^p$, $H_{\mu\nu}^p$ and $G_{\mu\nu}^p$ of SU($2$) subgroups are expressed in terms of electric potentials $A_{\mu}^p$, magnetic potentials $C_{\mu}^p$ and dual potentials $B_{\mu}^p$, respectively.

Using Eq.\;\eqref{s2eq11} and Eq.\;\eqref{g}, we can calculate restricted Lagrangian of G($2$),
\begin{align}\label{s2eq15}
\begin{aligned}
\mathcal{L}_{R}&=-\frac{1}{4}\hat{G}_{\mu\nu}^2=-\frac{1}{4}(G_{\mu\nu}^2+G_{\mu\nu}^{'2})=-\frac{1}{4}(F_{\mu\nu}^{2}\\
&+H_{\mu\nu}^{2}+2 F_{\mu\nu} H_{\mu\nu}+F_{\mu\nu}^{'2}+H_{\mu\nu}^{'2}+2 F'_{\mu\nu} H'_{\mu\nu})\\
&=-\frac{1}{2}\sum_{p=1}^{6} (\hat{G}_{\mu\nu}^p)^2=-\frac{1}{2}\sum_{p=1}^{6} (G_{\mu\nu}^p)^2\\
&=-\frac{1}{2}\sum_{p=1}^{6} (F_{\mu\nu}^p+H_{\mu\nu}^p)^2.
\end{aligned}
\end{align}
This restricted Lagrangian contains the non-Abelian monopole degrees of freedom explicitly, and hence, it is possible to study the dynamics of monopole gauge, independently.

In the beginning of this section, we have had a review on Cho decomposition of the SU($2$) gauge group. And by choosing the Abelian direction of this group in the $\hat{r}$ direction, we have found a pointlike monopole with Dirac string along the positive $z$ axis. Now, we use the method introduced by Cho to find G($2$) monopoles. SU($2$) contains one Abelian direction, but G($2$) contains two Abelian directions. To find monopoles of G($2$), we consider SU($2$) subgroups of G($2$) where each subgroup has only one Abelian direction. We recall that G($2$) can be covered by six SU($2$) subgroups. The color spaces of SU($2$) subgroups of G($2$) are the same as the color space of SU($2$) gauge group. Therefore, to find monopoles of SU($2$) subgroups, we can do the same discussions and calculations as what we have done for the SU($2$) gauge group. Monopoles of SU($2$) subgroups arise from the magnetic part of the field strength $\hat{G}_{\mu\nu}$ defined in Eq.\;\eqref{g} and Eq.\;\eqref{cg}. Hedgehog shapes for Abelian directions of SU($2$) subgroups are obtained by choosing $\hat{n}^p=\mid\hat{n}^p\mid\hat{r}$ and $C_{\mu}^p=\dfrac{1}{\mid\hat{n}^p\mid^2}\dfrac{1}{g}(1+\cos\theta)\partial_{\mu}\varphi$ as a magnetic potential for each SU($2$) subgroup. Components of these magnetic potentials are
\begin{equation}\label{ssg2} 
C_{\theta}^p=C_{r}^p=0,\quad C_{\varphi}^p=\dfrac{1}{\mid\hat{n}^p\mid^2}\dfrac{1}{g}\dfrac{(1+\cos\theta)}{r\sin\theta}.
\end{equation}
This describes a pointlike monopole at the origin with a Dirac string along the positive $z$ axis for each SU($2$) subgroup. Since the subgroups are Abelianized in $\hat{n}^p$ directions, near each monopole, the singularity of the gluon field is given by
\begin{equation}\label{ssg3} 
\hat{A}^{S}_p=\mathbf{C}^p T_{d}=\dfrac{1}{\mid\hat{n}^p\mid^2}\dfrac{1}{g}\dfrac{(1+\cos\theta)}{r\sin\theta}T_{d}\hat{\varphi},
\end{equation}
where $T_{d} (d=p=1,2,...,6)$ are diagonal generators of SU($2$) subgroups brought in Eq.\;\eqref{a2}. The summation on $p$ in Eq.\;\eqref{g} and the fact that G($2$) has two Abelian subgroups described by Eq.\;\eqref{e9} suggest that there exist only two independent monopoles for G($2$) gauge group. Therefore, the generalized quantization condition for G($2$) is labeled by two integers $N$ and $N'$,
\begin{align}\label{mc1}
\begin{aligned}
&exp\{-8\pi i g[\dfrac{1}{g}(N-\dfrac{N'}{2})T_3+\dfrac{1}{g}\dfrac{\sqrt{3}}{2}N'T_8]\}\\
&=exp\{(2\pi i) diag[-N,N-N',N',N,N'-N,-N',0]\}\\
&=1.
\end{aligned}
\end{align}
This leads to magnetic charges of G($2$),
\begin{equation}\label{mc2} 
g(N,N')=-\dfrac{8\pi}{g}[(N-\dfrac{N'}{2})T_3+\dfrac{\sqrt{3}}{2}N'T_8].
\end{equation}
It is clear from the above equation that there are six monopoles of minimal magnetic charges,
\begin{align}\label{mc3}
\begin{aligned}
&g(1,0)=g_1=-\dfrac{8\pi}{g}T_3,\\ &g(1,1)=g_2=-\dfrac{8\pi}{g}(\dfrac{1}{2}T_3+\dfrac{\sqrt{3}}{2}T_8),\\
&g(0,1)=g_3=-\dfrac{8\pi}{g}(-\dfrac{1}{2}T_3+\dfrac{\sqrt{3}}{2}T_8),\\ &g(2,1)=g_4=-3[\dfrac{8\pi}{g}(\dfrac{1}{2}T_3+\dfrac{\sqrt{3}}{6}T_8)],\\ &g(1,2)=g_5=-3[\dfrac{8\pi}{g}(\dfrac{\sqrt{3}}{3}T_8)],\\ 
&g(-1,1)=g_6=-3[\dfrac{8\pi}{g}(-\dfrac{1}{2}T_3+\dfrac{\sqrt{3}}{6}T_8)],
\end{aligned}
\end{align}
where only two of them are independent. Comparing the magnetic charges which can be obtained from Eq.\;\eqref{ssg3} with magnetic charges of Eq.\;\eqref{mc3}, it is confirmed that our choice for $C_{\mu}^p$ is suitable. We can express magnetic charges of Eq.  \eqref{mc3} in terms of positive roots of Fig.\;(\ref{rootdiagram}),
\begin{align}\label{eq14}
\begin{aligned}
g_p=-\dfrac{4\pi}{g{\mid\hat{n}^p\mid^2}}\mathbf{H}\cdot \mathbf{r}_p=\begin{cases}
-\dfrac{16\pi}{g}\mathbf{H}\cdot \mathbf{r}_p \quad &(p=1,2,3)\\
-3[\dfrac{16\pi}{g}\mathbf{H}\cdot \mathbf{r}_p]\quad &(p=4,5,6)
\end{cases},
\end{aligned}
\end{align}
where two diagonal generators of G($2$) are represented by a two-dimensional vector $\mathbf{H}$,
\begin{equation}\label{H}
\mathbf{H}=(T_3,T_8).
\end{equation}
Indeed, $\mathbf{H}\cdot \mathbf{r}_p$ are diagonal generators of SU($2$) subgroups \eqref{a2}. By comparing magnetic charges of Eq.\;\eqref{eq14} with roots of Fig.\;(\ref{rootdiagram}), we find out a beautiful concept behind the root diagram, which is a correspondence between the root vectors and magnetic charges. Magnetic charges $g_1$, $g_2$, and $g_3$ in Eq.\;\eqref{eq14} are equivalent to the magnetic charges of SU($3$) \cite{ripka}, as one of the subgroups of G($2$). They could easily be predicted by studying the roots $\mathbf{r}_1$, $\mathbf{r}_2$, and $\mathbf{r}_3$ of Fig.\;(\ref{rootdiagram}). These three roots form the root diagram of SU($3$) gauge group\cite{modern}.

In the next section, we briefly review Abelian gauge fixing method, and then we extract monopoles and their magnetic charges for G($2$) gauge group by this method.

\section{Abelian gauge fixing and magnetic monopoles of the G($2$)}\label{sec100}

In the Abelian gauge fixing, QCD is reduced into an Abelian gauge theory, where QCD monopoles appeared \cite{Suganuma,Ichie,pepe,ripka}. In this section, we find monopoles of G($2$) by this method. For non-Abelian gauge groups, Abelian gauge fixing is defined by  diagonalizing a suitable scalar field $\Phi (x)$. Hence, for G($2$) group,

\begin{equation}\label{e003}
\Omega (x)\Phi (x)\Omega^{\dagger} (x)=diag[\lambda_1 (x),\lambda_2 (x),...,\lambda_7 (x)],
\end{equation} 
where $\Omega (x)\in$ G($2$) is a gauge transformation function which represents a local rotation in color space of G($2$). $\lambda_i$'s are eigenvalues of diagonalized $\Phi (x)$, and there exist seven of them, since we work in seven-dimensional fundamental representation of G($2$). The scalar field $\Phi (x)$ is defined as the following: 
\begin{equation}\label{e001}
\Phi (x)=\Phi_a (x) T_a,
\end{equation}
where $T_a (a=1,2,...,14)$ are generators of G($2$) and $\Phi_a (x)$'s are the components of $\Phi$ in color space. Since the generators $T_a$ are traceless Hermitian matrices, the field $\Phi (x)=\Phi_a (x) T_a$ and also the diagonalized form of it which is given in Eq. \eqref{e003} are traceless matrices. The Abelian gauge fixing becomes ill-defined at some points in space. At these points, $\lambda_i (i=1,2,...,7)$ eigenvalues in Eq. \eqref{e003} become degenerate, the gluon field $A^{\mu}_{a}$ turns singular and magnetic monopoles appear. We investigate these degenerate points in G($2$) space and find monopoles and their associated monopole charges. 

For this purpose, we study degenerate points of the SU($2$) subspaces of G($2$).
In the vicinity of each one of the degeneracy points, there are at least two degenerate eigenvalues and the matrix $\Phi (x)$ is not diagonalized completely. By expressing the scalar field $\Phi (x)$ in terms of SU($2$) subgroups, we obtain six different forms for $\Phi (x)$ since G($2$) contains six SU($2$) subgroups brought in Eq.\eqref{a2}. We also find out that there are six possible degenerate cases since we can consider one degenerate case corresponds to each one of these scalar fields. For the first degenerate case which arises from $(1/{2} T_1,1/{2} T_2,1/{2} T_3)$ subgroup, the scalar field is written in the following form:
\vspace{-1.1cm}\begin{widetext}
\begin{align}\label{eq2}
\begin{aligned}
\Phi\simeq \dfrac{1}{8}\begin{pmatrix}
\varepsilon_3+\lambda&\varepsilon_1-i\varepsilon_2&&&&\multicolumn{2}{c}{\text{\kern-2.5em\smash{\raisebox{-3ex}{\mbox{\Huge 0}}}}}\\
\varepsilon_1+i\varepsilon_2&-\varepsilon_3+\lambda&&&&&\\
&&-2\lambda&&&&\\
&&&-\varepsilon_3-\lambda&-\varepsilon_1-i\varepsilon_2&&\\
&&&-\varepsilon_1+i\varepsilon_2&\varepsilon_3-\lambda&&\\
&&&&&2\lambda&\\
\multicolumn{2}{c}{\text{\kern-2em\smash{\raisebox{4ex}{\mbox{\Huge 0}}}}}&&&&&0\\
\end{pmatrix}=\Phi_8 t_8+\sum_{a=1}^{3} \Phi_a t_a,
\end{aligned}
\end{align}
\end{widetext}
where
\begin{equation}\label{phi}
\Phi_1=\varepsilon_1,\quad \Phi_2=\varepsilon_2,\quad \Phi_3=\varepsilon_3,\quad \Phi_8=\sqrt{3}\lambda,
\end{equation}
and we use the following appropriate definitions:
\begin{equation}\label{eq3}
t_1=\dfrac{1}{2}T_1,\quad t_2=\dfrac{1}{2}T_2,\quad t_3=\dfrac{1}{2}T_3,\quad t_8=\dfrac{1}{2}T_8.
\end{equation}
The eigenvalues of the scalar field \eqref{eq2} are almost degenerate and the first degenerate case is specified by
\begin{align}\label{eq1}
\begin{aligned}
&\lambda_1=\lambda_2=\dfrac{\lambda}{8},\quad\lambda_4=\lambda_5=-\dfrac{\lambda}{8},\\
&\lambda_3=-\lambda_6=-\dfrac{\lambda}{4},\quad\lambda_7=0,
\end{aligned}
\end{align}
where it is obtained when all components $\varepsilon_{a}\rightarrow 0$ as we approach to the degenerate points. Now, we need a gauge transformation $\Omega$ to transform the almost diagonal matrix in Eq.\;\eqref{eq2} into its diagonal form,
\vspace{-1.1cm}\begin{widetext}
\begin{align}\label{eq4}
\begin{aligned}
\Phi_8 t_8+\sum_{a=1}^{3} \Phi_a t_a \rightarrow  \Omega (\Phi_8 t_8+\sum_{a=1}^{3} \Phi_a t_a)\Omega^{\dagger}=\Phi_8 t_8+\varepsilon t_3=\Phi_8 t_8+\dfrac{1}{8}\begin{pmatrix}
\varepsilon&&&&&&\multicolumn{2}{c}{\text{\kern-2em\smash{\raisebox{-4ex}{\mbox{\Huge 0}}}}}\\
&-\varepsilon&&&&&\\
&&0&&&&\\
&&&-\varepsilon&&&\\
&&&&\varepsilon&&\\
&&&&&0&\\
\multicolumn{2}{c}{\text{\kern2em\smash{\raisebox{3ex}{\mbox{\Huge 0}}}}}&&&&&&0\\
\end{pmatrix},
\end{aligned}
\end{align}
\end{widetext}
where the eigenvalue $\varepsilon$ is
\begin{equation}\label{eq5}
\varepsilon=\sqrt{\Phi_1^2+\Phi_2^2+\Phi_3^2}.
\end{equation}
This shows that in the Abelian gauge of G($2$) group, the components of the scalar field are aligned along the diagonal generators $T_3$ and $T_8$ or combinations of them. In addition, $\Omega$ rotates the vector $\sum_{a=1}^{3} \Phi_a t_a$ to the Abelian direction $t_3$. The eigenvalues ,$\pm\varepsilon$, are degenerate at the points $\mathbf{r}_0$ in space where $\Phi_1(\mathbf{r}_0)=\Phi_2(\mathbf{r}_0)=\Phi_3(\mathbf{r}_0)=0$. Applying a Taylor expansion to $\Phi (\mathbf{r})$ near $\mathbf{r}_0$,
\begin{align}\label{eq6}
\begin{aligned}
\Phi (\mathbf{r})&=\Phi_8 (\mathbf{r}) t_8+\sum_{a=1}^{3} \Phi_a (\mathbf{r}) t_a=\Phi_8 (\mathbf{r}) t_8\\
&+\sum_{a=1}^{3} C_{ab}(x_b-x_{0b}) t_a,\quad C_{ab}=\dfrac{\partial \Phi_a}{\partial x_b} \mid_{\mathbf{r}=\mathbf{r_0}}.
\end{aligned}
\end{align}
The matrix $C_{ab}$ defines a coordinate system (denoted by prime) such that the field $\Phi (\mathbf{r'})$ in this coordinate system acquires a hedgehog shape,
\begin{align}\label{eq7}
\begin{aligned}
&\Phi (\mathbf{r'})=\Phi_8 (\mathbf{r'}) t_8+\sum_{a=1}^{3} x'_a t_a,\\
& x'_a=\sum_{b=1}^{3}C_{ab}(x_b-x_{0b}),
\end{aligned}
\end{align}
where the monopole appears from this hedgehog configuration. By working in the $x'$-coordinate frame and then omitting the primes, the degenerate point is placed at the origin. The topological natures are not changed when transformation between $x$ and $x'$ is applied. Thus, the degenerate points give monopoles in the Abelian gauge in terms of $x$-coordinate frame. In spherical coordinates, the scalar field is written in the following form:
\vspace{-1.1cm}\begin{widetext}
\begin{align}\label{eq8}
\begin{aligned}
\Phi (\mathbf{r})&=\Phi_8 (\mathbf{r}) t_8+ x_a t_a=\Phi_8 (\mathbf{r}) t_8+t_1 r \sin \theta \cos \varphi+t_2 r \sin \theta \sin \varphi+t_3 r \cos \theta\\
&=\Phi_8 (\mathbf{r}) t_8+\dfrac{r}{8}\begin{pmatrix}
\cos\theta&e^{-i\varphi}\sin\theta&&&&&&\multicolumn{2}{c}{\text{\kern-6em\smash{\raisebox{-3ex}{\mbox{\Huge 0}}}}}\\
e^{i\varphi}\sin\theta&-\cos\theta&&&&&\\
&&0&&&&\\
&&&-\cos\theta&-e^{i\varphi}\sin\theta&&\\
&&&-e^{-i\varphi}\sin\theta&\cos\theta&&\\
&&&&&0&\\
\multicolumn{2}{c}{\text{\kern2em\smash{\raisebox{3ex}{\mbox{\Huge 0}}}}}&&&&&&&0\\
\end{pmatrix}.
\end{aligned}
\end{align}
\end{widetext}
The matrix $\Omega$ which brings this matrix into a diagonal form is
\vspace{-1.1cm}\begin{widetext}
\begin{equation}\label{eq9}
\Omega (\theta,\varphi)=\begin{pmatrix}
e^{i\varphi}\cos\dfrac{\theta}{2}&\sin\dfrac{\theta}{2}&&&&&\multicolumn{2}{c}{\text{\kern-6em\smash{\raisebox{-3ex}{\mbox{\Huge 0}}}}}\\
-\sin\dfrac{\theta}{2}&e^{-i\varphi}\cos\dfrac{\theta}{2}&&&&&\\
&&1&&&&\\
&&&e^{-i\varphi}\cos\dfrac{\theta}{2}&\sin\dfrac{\theta}{2}&&\\
&&&-\sin\dfrac{\theta}{2}&e^{i\varphi}\cos\dfrac{\theta}{2}&&\\
&&&&&1&\\
\multicolumn{2}{c}{\text{\kern2em\smash{\raisebox{3ex}{\mbox{\Huge 0}}}}}&&&&&&1\\
\end{pmatrix}.
\end{equation}
\end{widetext}
Under this gauge transformation, the gauge field is transformed as
\begin{equation}\label{eq10}
\mathbf{A}=\mathbf{A}_a T_a\rightarrow \Omega(\mathbf{A}+\dfrac{1}{ig}\mathbf{\nabla})\Omega^{\dagger}.
\end{equation}
Since the original gauge field $\mathbf{A}$ is regular, the first term ,$\Omega\mathbf{A}\Omega^{\dagger}$, is also regular. The second term becomes singular where by considering the gradient in spherical coordinates it is given by
\begin{align}\label{eq11}
\begin{aligned}
\dfrac{1}{ig}\Omega\mathbf{\nabla}\Omega^{\dagger}&=\dfrac{4}{g}[-\hat{\theta}e^{i\varphi}t_2-\hat{\varphi}\dfrac{1+\cos \theta}{r \sin \theta}t_3\\
&+\hat{\varphi}\dfrac{1}{r}(\cos \varphi t_1-\sin \varphi t_2)].
\end{aligned}
\end{align}
All terms in Eq.\;\eqref{eq11} are regular except the second term which obtains a singularity when $\theta\rightarrow0$  on the positive z axis. In the Abelian gauge, the gluon field $\mathbf{A}$ can be separated into a regular $\mathbf{A}^R$ and a singular $\mathbf{A}^S$ part,
\begin{equation}\label{eq12}
\mathbf{A}=\mathbf{A}_a T_a=\mathbf{A}^R_o T_o+\mathbf{A}^S_d T_d=\mathbf{A}^R_o T_o-\dfrac{4}{g}(\hat{\varphi}\dfrac{1+\cos \theta}{r \sin \theta}t_3),
\end{equation}
where ``$a$", ``$d$", ``$o$" denote summations on all generators, diagonal generators and off diagonal generators of G($2$), respectively. Only the diagonal part of the gauge field acquires a singular form. Thus, in the vicinity of the singular point of the first degenerate case \eqref{eq1}, the diagonal gluon $\mathbf{A}_3$ feels the presence of a monopole with the following magnetic charge:
\begin{equation}\label{eq13}
g_1=-\dfrac{16\pi}{g}t_3=-\dfrac{16\pi}{g}(\dfrac{T_3}{2}).
\end{equation}

As another example, we study the fourth degenerate case which is defined by
\begin{align}\label{z1}
\begin{aligned}
&\lambda_2=\lambda_6=\dfrac{\lambda}{24},\quad\lambda_3=\lambda_5=-\dfrac{\lambda}{24},\\
&\lambda_1=\lambda_4=\lambda_7=0,
\end{aligned}
\end{align}
where it arises from the fourth SU($2$) subgroup in Eq.\;\eqref{a2}. In the vicinity of this degeneracy point, the scalar field is written in an almost diagonal form,
\vspace{-1.1cm}\begin{widetext}
\begin{align}\label{z2}
\begin{aligned}
\Phi\simeq \dfrac{1}{24}\left(\begin{matrix}
2\varepsilon_3&0&0&0&0&0&\sqrt{2}(\varepsilon_1-i\varepsilon_2)\\
0&-\varepsilon_3+\lambda&0&0&0&\varepsilon_1+i\varepsilon_2&0\\
0&0&-\varepsilon_3-\lambda&0&-\varepsilon_1-i\varepsilon_2&0&0\\
0&0&0&-2\varepsilon_3&0&0&\sqrt{2}(\varepsilon_1+i\varepsilon_2)\\
0&0&-\varepsilon_1+i\varepsilon_2&0&\varepsilon_3-\lambda&0&0\\
0&\varepsilon_1-i\varepsilon_2&0&0&0&\varepsilon_3+\lambda&0\\
\sqrt{2}(\varepsilon_1+i\varepsilon_2)&0&0&\sqrt{2}(\varepsilon_1-i\varepsilon_2)&0&0&0\\
\end{matrix}\right)=\Phi_8 t_8+\sum_{a=1}^{3} \Phi_a t_a,
\end{aligned}
\end{align}
\end{widetext}
where
\begin{align}\label{z3}
\begin{aligned}
&\Phi_1=\varepsilon_1,\quad \Phi_2=\varepsilon_2,\quad \Phi_3=\varepsilon_3,\quad \Phi_8=\dfrac{\lambda}{3},\\
&t_1=\dfrac{1}{2\sqrt{3}}T_9,\quad t_2=\dfrac{1}{2\sqrt{3}}T_{10},\quad t_3=\dfrac{1}{4}T_3+\dfrac{\sqrt{3}}{12}T_8,\\ &t_8=-\dfrac{1}{4}T_3+\dfrac{\sqrt{3}}{4}T_8.
\end{aligned}
\end{align}
Following the same discussions we have done for the previous SU($2$) subgroup in Eqs.\;{\eqref{eq4} to \eqref{eq7}}, we find out that the scalar field in Eq.\;\eqref{z2} acquires a hedgehog shape,
\vspace{-1.1cm}\begin{widetext}
\begin{align}\label{z4}
\begin{aligned}
\Phi (\mathbf{r})&=\Phi_8 (\mathbf{r}) t_8+ x_a t_a=\Phi_8 (\mathbf{r}) t_8+t_1 r \sin \theta \cos \varphi+t_2 r \sin \theta \sin \varphi+t_3 r \cos \theta\\
&=\Phi_8 (\mathbf{r}) t_8+\dfrac{r}{24}\left(\begin{matrix}
2\cos\theta&0&0&0&0&0&\sqrt{2}\sin\theta e^{-i\varphi}\\
0&-\cos\theta&0&0&0&\sin\theta e^{i\varphi}&0\\
0&0&-\cos\theta&0&-\sin\theta e^{i\varphi}&0&0\\
0&0&0&-2\cos\theta&0&0&\sqrt{2}\sin\theta e^{i\varphi}\\
0&0&-\sin\theta e^{-i\varphi}&0&\cos\theta&0&0\\
0&\sin\theta e^{-i\varphi}&0&0&0&\cos\theta&0\\
\sqrt{2}\sin\theta e^{i\varphi}&0&0&\sqrt{2}\sin\theta e^{-i\varphi}&0&0&0\\
\end{matrix}\right).
\end{aligned}
\end{align}
\end{widetext}
The matrix $\Omega$ which brings this matrix into a diagonal form is
\vspace{-1.1cm}\begin{widetext}
 \begin{align}\label{z5}
 \begin{aligned}
 \Omega(\theta,\varphi)=\left(\begin{matrix}
 e^{2i\varphi}\cos^{2}\dfrac{\theta}{2}&0&0&\sin^{2}\dfrac{\theta}{2}&0&0&\dfrac{1}{\sqrt{2}}e^{i\varphi}\sin\theta\\
 0&e^{-i\varphi}\cos\dfrac{\theta}{2}&0&0&0&-\sin\dfrac{\theta}{2}&0\\
 0&0&e^{-i\varphi}\cos\dfrac{\theta}{2}&0&\sin\dfrac{\theta}{2}&0&0\\
 -\sin^{2}\dfrac{\theta}{2}&0&0&-e^{-2i\varphi}\cos^{2}\dfrac{\theta}{2}&0&0&\dfrac{1}{\sqrt{2}}e^{-i\varphi}\sin\theta\\
 0&0&\sin\dfrac{\theta}{2}&0&-e^{i\varphi}\cos\dfrac{\theta}{2}&0&0\\
 0&\sin\dfrac{\theta}{2}&0&0&0&e^{i\varphi}\cos\dfrac{\theta}{2}&0\\
 -\dfrac{1}{\sqrt{2}}e^{i\varphi}\sin\theta&0&0&\dfrac{1}{\sqrt{2}}e^{-i\varphi}\sin\theta&0&0&\cos\theta\\
 \end{matrix}\right).
 \end{aligned}
 \end{align}
\end{widetext}
By this gauge transformation, the singular term of the transformed gauge field becomes
\begin{equation}\label{z6}
[\dfrac{1}{ig}\Omega\nabla\Omega^{\dagger}]_{singular}=-\dfrac{12}{g}[\hat{\varphi}\dfrac{1+\cos\theta}{r\sin\theta}(\dfrac{1}{4}T_3+\dfrac{\sqrt{3}}{12}T_8)].
\end{equation}
Therefore a monopole with a Dirac string along the positive z axis is formed. The magnetic charge of this monopole is
 \begin{equation}\label{z7} 
 g_4=-\dfrac{4\pi}{g}(3 T_3+\sqrt{3} T_8).
 \end{equation}
Studying other degeneracies of the eigenvalues, one can obtain other pointlike monopoles of G($2$). The calculation method is the same as what we have done for the first and the fourth cases. Thus, by applying Abelian gauge fixing method for G($2$), six monopoles with magnetic charges given in Eq.\;\eqref{eq14} is observed. However, as we have already discussed, only two of those six monopoles are independent. This result is in agreement with what we have obtained by the Cho decomposition method. 

\section{Conclusion}\label{sec20}

In this paper, monopoles of the G($2$) gauge group as a smallest simply connected gauge group are studied. Two well-known methods: the Cho decomposition and Abelian gauge fixing are applied to extract these topological defects existing in the low energy limit of QCD. Using both the Cho decomposition and Abelian gauge fixing methods, monopoles of G($2$) are derived, and it is shown that the results of the two methods are the same. Generators of G($2$) which are obtained in some previous papers are rearranged in such a way that the SU($2$) and SU($3$) decomposition contributions of G($2$) fundamental representation are observed more easily. 

\section*{Acknowledgement}

We are grateful to the research council of University of Tehran for supporting this study.

\appendix
\section{G($2$) generators} \label{Calcu}

\setcounter{equation}{0}
 \renewcommand{\theequation}{\Alph{section}.\arabic{equation}}
 
In this Appendix, the generators of G($2$) fundamental representation are represented. In the papers \cite{pepe1,modern}, the generators of this representation are found as the following:
  \begin{align}\label{g2generators}
  \begin{aligned}
  E_{1}&=\dfrac{1}{2\sqrt{2}}(\rvert 1\rangle\langle 2\rvert-\rvert 5\rangle\langle 4\rvert),\\
  E_{2}&=\dfrac{1}{2\sqrt{2}}(\rvert 1\rangle\langle 3\rvert-\rvert 6\rangle\langle 4\rvert),\\
  E_{3}&=\dfrac{1}{2\sqrt{2}}(\rvert 2\rangle\langle 3\rvert-\rvert 6\rangle\langle 5\rvert),\\
  E_{4}&=\dfrac{1}{2\sqrt{6}}(\rvert 6\rangle\langle 2\rvert-\rvert 5\rangle\langle 3\rvert+\sqrt{2}\rvert 1\rangle\langle 7\rvert+\sqrt{2}\rvert 7\rangle\langle 4\rvert),\\
  E_{5}&=\dfrac{1}{2\sqrt{6}}(\rvert 1\rangle\langle 5\rvert-\rvert 2\rangle\langle 4\rvert+\sqrt{2}\rvert 6\rangle\langle 7\rvert+\sqrt{2}\rvert 7\rangle\langle 3\rvert),\\
  E_{6}&=\dfrac{1}{2\sqrt{6}}(\rvert 4\rangle\langle 3\rvert-\rvert 6\rangle\langle 1\rvert+\sqrt{2}\rvert 2\rangle\langle 7\rvert+\sqrt{2}\rvert 7\rangle\langle 5\rvert),\\
  H_1&=\dfrac{1}{4}(\rvert 1\rangle\langle 1\rvert-\rvert 2\rangle\langle 2\rvert-\rvert 4\rangle\langle 4\rvert+\rvert 5\rangle\langle 5\rvert),\\
  H_2&=\dfrac{1}{4\sqrt{3}}(\rvert 1\rangle\langle 1\rvert+\rvert 2\rangle\langle 2\rvert-2\rvert 3\rangle\langle 3\rvert-\rvert 4\rangle\langle 4\rvert-\rvert 5\rangle\langle 5\rvert\\
  &+2\rvert 6\rangle\langle 6\rvert),
  \end{aligned}
  \end{align}
where $\rvert\mathbf{m}_{a}\rangle (a=1,2,...,7)$ stand for the eigenstates of the fundamental representation of G($2$) obtained in Eq.\;\eqref{hk}. The above stepping generators are real, but we would like the eight generators of G($2$) to have the complex form of Eq.\;\eqref{e5}, such that the SU($3$) decomposition of G($2$) is explicitly observed. For this purpose, we use the $T_i (i=1,2,...,14)$ rather than $H_k$ and $E_{\alpha}$ as generators of the fundamental representation of G($2$). These generators are obtained from the the following equations: 
  \begin{align}\label{ap5}
  \begin{aligned}
  &E_{\pm1}=\dfrac{1}{\sqrt{2}}(T_1\pm iT_2),\quad E_{\pm2}=\dfrac{1}{\sqrt{2}}(T_4\pm iT_5),\\ &E_{\pm3}=\dfrac{1}{\sqrt{2}}(T_6\pm iT_7), \quad
  E_{\pm4}=\dfrac{1}{\sqrt{2}}(T_{9}\pm iT_{10}),\\ &E_{\pm5}=\dfrac{1}{\sqrt{2}}(T_{11}\pm iT_{12}), \quad
  E_{\pm6}=\dfrac{1}{\sqrt{2}}(T_{13}\pm iT_{14}),\\
  &H_1=T_3,\quad H_2=T_8.
  \end{aligned}
  \end{align}
The normalization condition is defined as $tr (T_i T_j)=1/4\delta_{ij}$. From this form of $T_i$'s, one can learn many interesting properties of SU($2$) and SU($3$) subgroups of G($2$). With this representation, it is obvious that the first eight generators of G($2$) contain SU($3$) generators, $\lambda_i (i=1,2,...,8)$, as presented in Eq.\;\eqref{e5}. In addition, the decomposition of G($2$) to its SU($3$) subgroup brought in Eq.\;\eqref{e4} is more understandable.

The space of G($2$) can be covered with six SU($2$) subgroups,
\vspace{-1.1cm}\begin{widetext}
\begin{align}\label{a2}
\begin{aligned}
& (\dfrac{1}{2}T_1,\dfrac{1}{2}T_2,\dfrac{1}{2}T_3),\quad
(\dfrac{1}{2}T_4,\dfrac{1}{2}T_5,\dfrac{1}{4}T_3+\dfrac{\sqrt{3}}{4}T_8),\quad
(\dfrac{1}{2}T_6,\dfrac{1}{2}T_7,-\dfrac{1}{4}T_3+\dfrac{\sqrt{3}}{4}T_8),\\
&(\dfrac{1}{2\sqrt{3}}T_9,\dfrac{1}{2\sqrt{3}}T_{10},\dfrac{1}{4}T_3+\dfrac{\sqrt{3}}{12}T_8),\quad
(\dfrac{1}{2\sqrt{3}}T_{11},\dfrac{1}{2\sqrt{3}}T_{12},\dfrac{\sqrt{3}}{6}T_8),\quad
(\dfrac{1}{2\sqrt{3}}T_{13},\dfrac{1}{2\sqrt{3}}T_{14},-\dfrac{1}{4}T_3+\dfrac{\sqrt{3}}{12}T_8),
\end{aligned}
\end{align}
\end{widetext}
where each category contains one diagonal generator and two off diagonal generators, the same as SU($2$) gauge group.

\section*{}


\begin{thebibliography}{9}

\bibitem{confinement1}
G. Cossu, M. D'Elia, A. D. Giacomo, B. Lucini, and C. Pica, J. High Energy Phys. 10 (2007) 100.

\bibitem{confinement2}
L. Liptak and S. Olejnik, Phys. Rev. D \textbf{78}, 074501 (2008).

\bibitem{well1}
B. H. Wellegehausen, A. Wipf, and C. Wozar, Phys. Rev. D \textbf{83}, 016001 (2011).

\bibitem{well2}
B. H. Wellegehausen, A. Wipf, and C. Wozar, Phys. Rev. D \textbf{83}, 114502 (2011).

\bibitem{confinement3}
C. Bonati, J. High Energy Phys. 03 (2015) 006.

\bibitem{vortex1}
J. Greensite, K. Langfeld, S. Olejnik, H. Reinhardt, and T. Tok, Phys. Rev. D \textbf{75}, 034501 (2007).

\bibitem{deldar1}
S. Deldar, H. Lookzadeh, and S. M. Hosseini Nejad, Phys. Rev. D \textbf{85}, 054501 (2012).

\bibitem{deldar2}	
S. M. Hosseini Nejad and S. Deldar,  Phys. Rev. D \textbf{89}, 014510 (2014).

\bibitem{deldar3}
S. M. Hosseini Nejad and S. Deldar, J. High Energy Phys. 03 (2015) 016.

\bibitem{shnir}
Ya. Shnir and G. Zhilin, Phys. Rev. D \textbf{92}, 045025 (2015).

\bibitem{kondo}
R. Matsudo and K-I. Kondo, Phys. Rev. D \textbf{94}, 045004 (2016).

\bibitem{fn1}
L. Faddeev and A. J. Niemi, Phys. Rev. Lett. \textbf{82}, 1624 (1999).

\bibitem{fn2}
L. Faddeev and A. J. Niemi, Phys. Lett. B \textbf{464}, 90 (1999).

\bibitem{fn3}
L. Faddeev and A. J. Niemi, Phys. Lett. B \textbf{525}, 195 (2002).

\bibitem{chord211980}
Y. M. Cho, Phys. Rev. D. \textbf{21}, 1080 (1980).

\bibitem{chord231981}
Y. M. Cho, Phys. Rev. D. \textbf{23}, 2415 (1981).

\bibitem{Suganuma}
H. Suganuma, S. Sasaki, and H. Toki, Nuclear. Phys. B \textbf{435}, 207 (1995).

\bibitem{Ichie}
H. Ichie  and H. Suganuma, arXiv: hep-lat/9906005v1.

\bibitem{pepe}
Ph. de Forcrand and M. Pepe, Nucl. Phys. B \textbf{598}, 557 (2001).

\bibitem{ripka}
G. Ripka, Lect. Notes Phys. \textbf{639}, 1(2004).

\bibitem{pepe1}
K. Holland, P. Minkowski, M. Pepe, and U. J. Wiese, Nucl. Phys. B \textbf{668}, 207 (2003).

\bibitem{pepe2}
M. Pepe and U. J. Wiese, Nucl. Phys. B \textbf{768}, 21 (2007).

\bibitem{instanton1}
M. Gunaydin and H. Nicolai, Phys. Lett. B \textbf{351}, 169 (1995).

\bibitem{instanton2}
I. Bauer, T. A. Ivanova, O. Lechtenfeld, and F. Lubbe, J. High Energy Phys. 10 (2010) 044.

\bibitem{instanton3}
A. S. Haupt, T. A. Ivanova, O. Lechtenfeld, and A. D. Popov, Phys. Rev. D \textbf{83}, 105028 (2011).

\bibitem{modern}
R. E. Behrends, J. Dreitlein, C. Fronsdal, and W. Lee, Rev. Mod. Phys. \textbf{34}, 1 (1962).

\bibitem{grouptheory1}
G. Costa and G. Fogli, \textit{Symmetries and Group
Theory in Particle Physics, An Introduction to Space-time and Internal
Symmetries}, Lecture Notes in Physics (Springer, Berlin, 2012), Vol. 823.

\bibitem{oxman2008}
L. E. Oxman, J. High Energy Phys. 12 (2008) 089.

\bibitem{cholett1980}
Y. M. Cho, Phys. Rev. Lett. \textbf{44}, 1115 (1980).

\bibitem{cho2014}
Y. M. Cho, Int. J. Mod. Phys. A. \textbf{29}, 1450013 (2014).




\end{thebibliography}
\end{document}